%% file: ttjj_TOP23.tex
\documentclass[12pt]{article}
\usepackage{graphicx}
\usepackage{mathtools}
\usepackage{subcaption}
\usepackage{cite}
\usepackage{hyperref}

\newcommand\pubnumber{MS-TP-24-04}
\newcommand\pubdate{January 15, 2024}

\def\institute{Institute for Theoretical Particle Physics and Cosmology,\\
RWTH Aachen University, D-52056 Aachen, Germany\\
~\\
Institute of Theoretical Physics\\
University of Münster, D-48149 Münster, Germany}
\def\authemail{\footnote{Contact: michele.lupattelli@uni-muenster.de}}

\def\Title#1{\begin{center} {\Large #1 } \end{center}}
\def\Author#1{\begin{center}{ \sc #1} \end{center}}
\def\Address#1{\begin{center}{ \it #1} \end{center}}

\newcommand\pubblock{\rightline{\begin{tabular}{l} \pubnumber\\
         \pubdate  \end{tabular}}}
\newenvironment{Abstract}{\begin{quotation}  }{\end{quotation}}
\newenvironment{Presented}{\begin{quotation} \begin{center} 
             PRESENTED AT\end{center}\bigskip 
      \begin{center}\begin{large}}{\end{large}\end{center} \end{quotation}}
\def\Acknowledgements{\bigskip  \bigskip \begin{center} \begin{large}
             \bf ACKNOWLEDGEMENTS \end{large}\end{center}}

\input econfmacros.tex

\begin{document}
\begin{titlepage}
\pubblock

\vfill
\Title{$t\bar{t}jj$ at the LHC: a study on the additional jets}
\vfill
\Author{Michele Lupattelli\authemail}
\Address{\institute}
\vfill
\begin{Abstract}
In this contribution we report on a recent calculation of the $pp\rightarrow t\bar{t}jj$ process
performed in the dileptonic decay channel using the narrow-width approximation.
The next-to-leading order corrections in QCD are included both in the production and decay of the top quarks.
For the first time, the additional light jet activity is consistently included at the matrix-element level not only
in the production stage of the top-quark pair but also in the decays.
The size of the next-to-leading order corrections and the impact of these additional contributions are investigated.
\end{Abstract}
\vfill
\begin{Presented}
$16^\mathrm{th}$ International Workshop on Top Quark Physics\\
(Top2023), 24--29 September, 2023
\end{Presented}
\vfill
\end{titlepage}
\def\thefootnote{\fnsymbol{footnote}}
\setcounter{footnote}{0}

\section{Introduction}

One of the most important processes studied at the LHC is the production of a Higgs boson
in association with a top-quark pair $t\bar{t}H$.
Its significance comes from the fact that it allows to directly probe the
top-Yukawa coupling.
However, this is a very rare process, representing only $1\%$ of the total Higgs boson
production rate~\cite{LHCHiggsCrossSectionWorkingGroup:2016ypw}.
Furthermore, the Higgs boson (as well as the top quark) is an unstable particle that, in
the experimental analyses, needs to be reconstructed from its decay products.

The Higgs-boson decay channel with the largest branching ratio ($58\%$) is $H \rightarrow b\bar{b}$~\cite{LHCHiggsCrossSectionWorkingGroup:2016ypw,CMS:2018hnq,ATLAS:2017fak,ATLAS:2021qou}.
Thus, one would naively think that this is the best channel to investigate this process.
However, it turns out that channels with smaller branching ratio, such as $H \rightarrow \gamma \gamma$~\cite{CMS:2018uxb,ATLAS:2018mme},
are considerably better at reconstructing the signal process, because the systematic uncertainties
are substantially smaller.
The problem of the $H \rightarrow b\bar{b}$ decay channel for this process is that the
final state presents at least four $b$ jets (two produced in the Higgs-boson decay and two
in the top-quark and antiquark decays).
The many possible combinations of $b$ jets make the reconstruction of the unstable
particles complicated.
In other words, the signal process represents a background to itself, often referred to
as \textit{combinatorial background}.

Adding to the complexity, the LHC is a top-quark factory, and top-quark pairs are often
produced with additional jets~\cite{ATLAS:2016qjg,ATLAS:2018acq}.
To properly reconstruct $t\bar{t}H(H\rightarrow b\bar{b})$, all the backgrounds need to be
kept under control.
The $pp \rightarrow t\bar{t}b\bar{b}$ process is the main background to the signal
process, since has the very same final state.
For that reason, it is referred to as \textit{irreducible background}.
Moreover, since the $b$-jet tagging efficiency in experiments is not perfect, light jets can
be misidentified as $b$ jets and enter the analysis.
Therefore, also the $pp \rightarrow t\bar{t}jj$ process needs to be taken into account, and is referred
to as \textit{reducible background}.
Experimental measurements for these background processes can be found in Ref.~\cite{ATLAS:2018fwl,CMS:2020grm}.

Theoretical studies play a crucial role in keeping these backgrounds under control.
Several calculations of the $t\bar{t}b\bar{b}$ process with next-to-leading order (NLO)
accuracy in QCD have been carried out, initially treating the top quark as a stable
particle~\cite{Bevilacqua:2009zn,Bredenstein:2008zb,Bredenstein:2009aj,Bredenstein:2010rs,Worek:2011rd,Bevilacqua:2014qfa,Kardos:2013vxa,Cascioli:2013era}, and later including the top-quark decays within
the parton shower~\cite{Garzelli:2014aba,Bevilacqua:2017cru}
or in the matrix element at leading-order (LO)~\cite{Jezo:2018yaf}.
The latest predictions have been obtained with fixed-order accuracy in the dileptonic
decay channel of the top quarks, both including full off-shell effects~\cite{Bevilacqua:2021cit,Denner:2020orv} and in the
narrow-width approximation (NWA)~\cite{Bevilacqua:2022twl}.
These two approaches preserve spin correlations and allow for a treatment of top-quark
decays beyond LO.
The NLO QCD corrections for this process are large and reduce the theoretical uncertainty,
which, however, is still quite sizable at NLO.
The full off-shell effects turn out to be relevant for observables with kinematic edges.
The $b$-quark content of the process has been extensively studied and, in Ref.~\cite{Bevilacqua:2022twl}, the
origin of the $b$ jets has been investigated, providing a theoretical prescription
to distinguish the $b$ jets emitted by the top quarks in their decay from the $b$ jets
produced in gluon splittings.
Moreover, in Ref.~\cite{Bevilacqua:2021cit} a prediction for the integrated fiducial cross section has been
compared to the measurement performed by the ATLAS collaboration~\cite{ATLAS:2018fwl}.
The two results are in very good agreement with each other, being only $0.6\sigma$ apart.

Similarly, the first NLO QCD predictions for $t\bar{t}jj$ have been obtained for stable
top quarks~\cite{Bevilacqua:2010ve,Bevilacqua:2011aa}.
Then, top-quark decays have been included at LO, and the fixed-order predictions have
been matched to a parton shower~\cite{Hoeche:2014qda,Hoche:2016elu,Gutschow:2018tuk}.
In these kind of calculations, at the matrix-element level the additional light jets are
present only in the production stage of the top-quark pair.
The additional light jet activity in the decays as well as the corrections
to the top-quark decays are only approximated by the parton shower.
This is obtained by merging $t\bar{t}+0,1,2$ jets samples and matching
the merged sample to the parton shower.
Recently, these contributions have been computed exactly at the matrix-element level for the
first time~\cite{Bevilacqua:2022ozv}.
Specifically, the dileptonic decay channel
$p p \rightarrow \ell^+ \nu_e \ell^- \bar{\nu}_\mu b \bar{b} j j + X$
has been studied in the NWA for LHC center-of-mass energy of $13$ TeV.
This contribution reports on this calculation.

\section{The $pp\rightarrow t\bar{t}jj$ process in QCD}
In Ref.~\cite{Bevilacqua:2022ozv} the $p p \rightarrow \ell^+ \nu_e \ell^- \bar{\nu}_\mu b \bar{b} j j + X$
process has been studied at NLO QCD accuracy in the NWA.
The NWA applies when an unstable particle has a mass which is considerably larger than
its decay width.
In this situation, the propagator factor can be approximated by a Dirac delta
\begin{equation}
\frac{1}{(p^2-m^2)^2 + m^2\Gamma^2} \rightarrow \frac{\pi}{m \Gamma} \delta(p^2 - m^2),
\end{equation}
which forces the on-shell production of the particle.
This approximation neglects $\mathcal{O}(\Gamma/m)$ effects~\cite{Uhlemann:2008pm} which, for the process at
hand, are of the order of $0.8\%$ at the integrated level.
This effects can be enhanced in the high-energy tails of distributions or in
correspondence of kinematic edges.
Most importantly, the NWA preserves top-quark spin correlations and allows for a treatment
of top-quark decays beyond LO.
The NWA induces a factorization of the cross section into a part that describes the on-shell
production of the heavy resonances and a part that describes the subsequent decays.
This factorization considerably simplifies the calculation, as only double-resonant diagrams
are considered.
The calculation has been carried out with the Monte-Carlo framework \textsc{Helac-Nlo}~\cite{Bevilacqua:2011xh}.
The theoretical predictions have been stored in the form of modified Les Houches event files~\cite{Alwall:2006yp}
and ROOT ntuples~\cite{Bern:2013zja,Antcheva:2009zz}.
They allow to obtain brand new predictions for the process with a different setup,
i.e. that employ different (more exclusive) kinematic cuts, renormalization and factorization
scales, PDF sets, without the need to run again from scratch such a complicated calculation.
They also allow to define new observables.

\begin{figure}[]
\centering
\begin{subfigure}{0.3\textwidth}
  \includegraphics[trim={0 15cm 0 0}, clip, width=\linewidth]{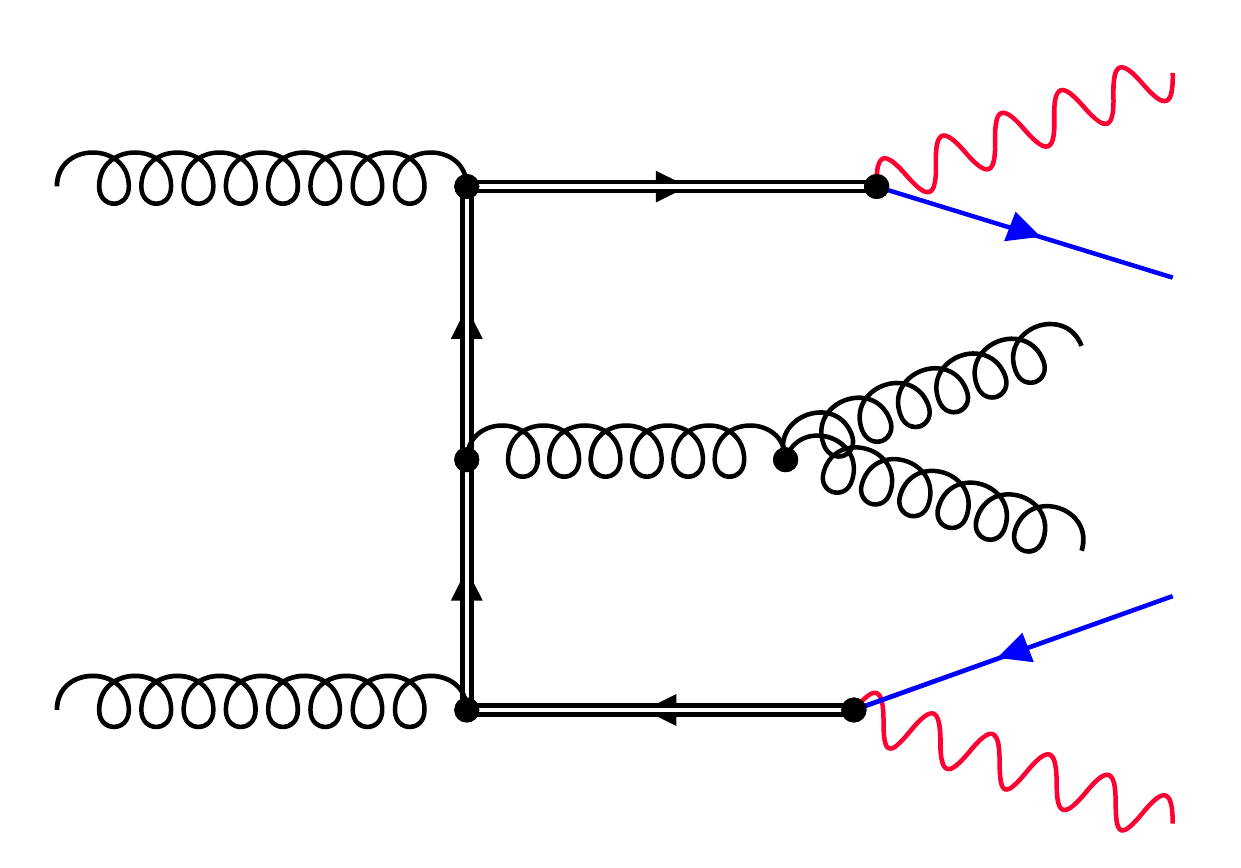}
  \caption*{\textit{Prod.}}
\end{subfigure}
\hfill
\begin{subfigure}{0.32\textwidth}
  \includegraphics[trim={0 14cm 0 0}, clip, width=\linewidth]{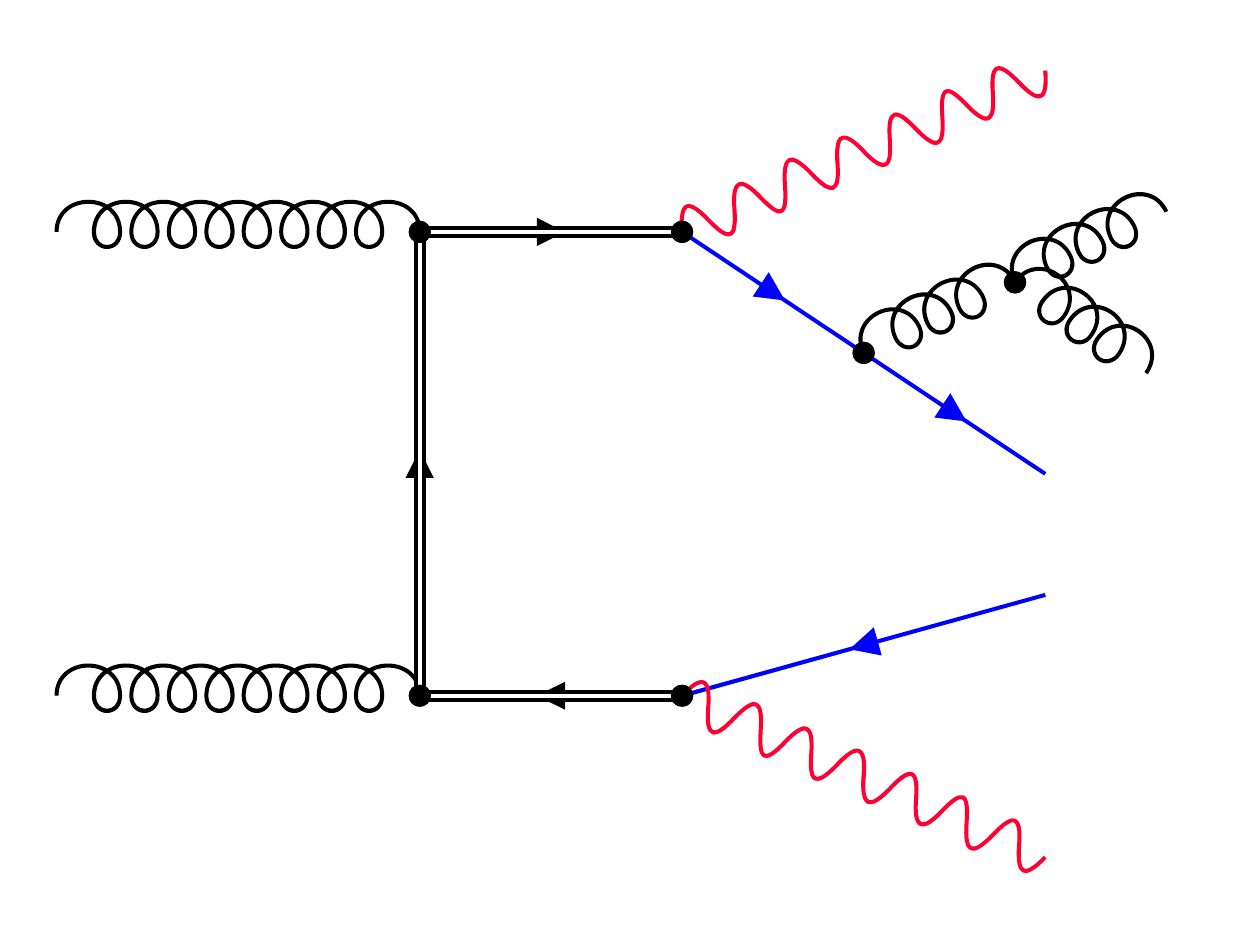}
  \caption*{\textit{Decay}}
\end{subfigure}
\hfill
\begin{subfigure}{0.3\textwidth}
  \includegraphics[trim={0 13cm 0 0}, clip, width=\linewidth]{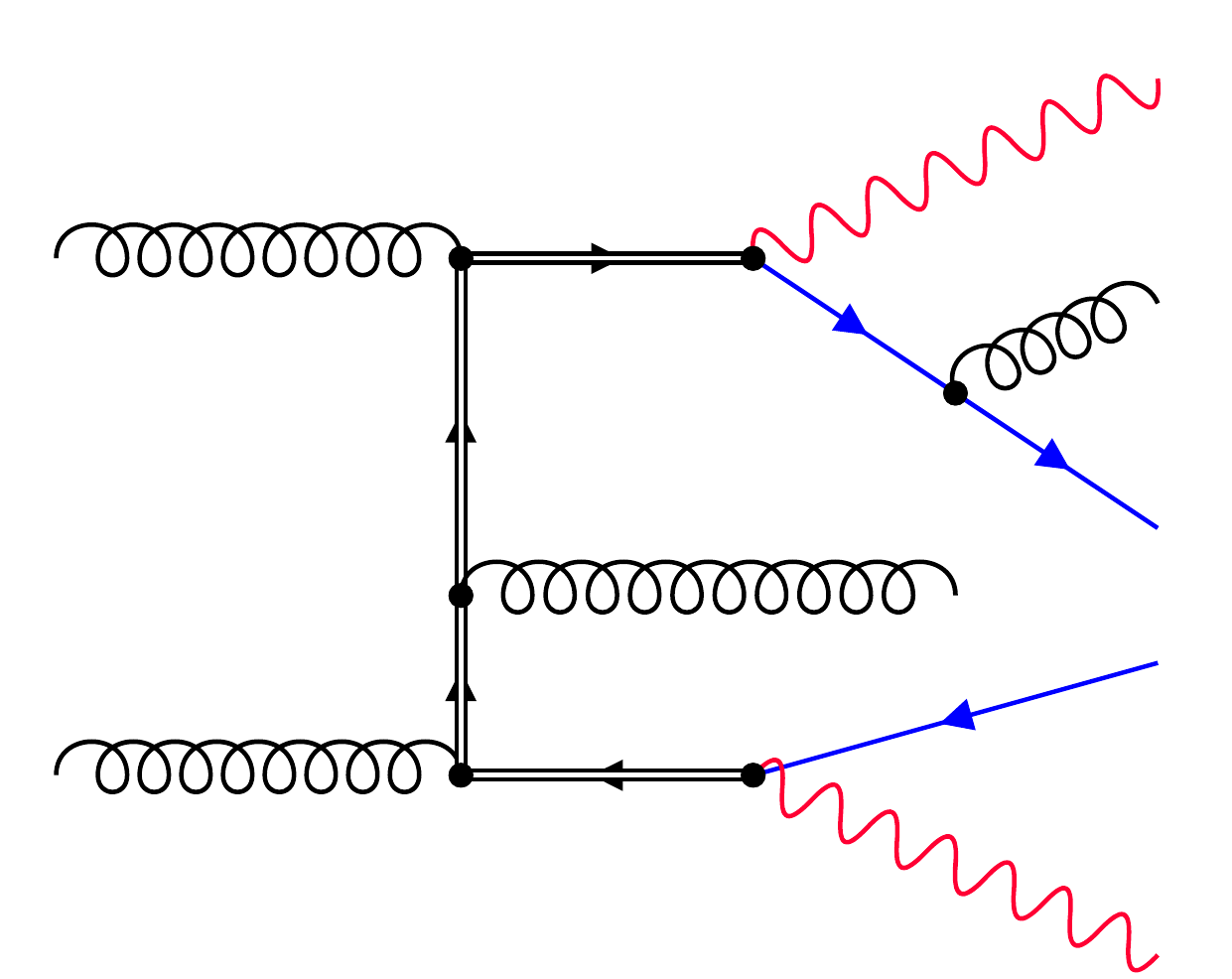}
  \caption*{\textit{Mix}}
\end{subfigure}
\caption{\textit{Various kind of contributions to the NWA $t\bar{t}jj$ cross section at LO. The black thick lines represent top (anti)quarks, the blue lines the bottom (anti)quarks and the red lines the W bosons. Figure taken from Ref.~\cite{Lupattelli:2023fgp}.}}
\label{fig:ttjj_LO_FD}
\end{figure}

The factorization of the cross section in production times decays induced by the NWA
allows to distinguish three main contributions to the cross section, depending on where
the additional jet activity occurs.
At LO, the cross section can be expressed
\begin{equation}
d\sigma^{\text{LO}}_{\text{Full}} = \bigl( \Gamma_{t,\text{NWA}}^{\text{LO}}\bigr )^{-2} \bigl (
\overbrace{ 
d\sigma^{\text{LO}}_{t\bar{t}jj} d\Gamma_{t\bar{t}}^{\text{LO}}}^{\text{Prod.}} + 
\overbrace{
d\sigma^{\text{LO}}_{t\bar{t}} d\Gamma_{t\bar{t}jj}^{\text{LO}} }^{\text{Decay}} +
\\ 
\overbrace{
d\sigma^{\text{LO}}_{t\bar{t}j} d\Gamma_{t\bar{t}j}^{\text{LO}} }^{\text{Mix}}
\bigr ),
\end{equation}
where $\Gamma_{t,\text{NWA}}^{\text{LO}}$ is the LO top-quark decay width computed in the NWA limit.
The first term includes contributions where the additional jet activity occurs exclusively
during the production stage of the top-quark pair and, as such, is labeled as \textit{Prod.}.
The second term incorporates contributions where the additional jets are emitted only
in the decay stage of the top-quark pair and, as such, is labeled as \textit{Decay}.
Finally, the \textit{Mix} term includes the contributions where light jets are present both in
production and decay.
Fig.~\ref{fig:ttjj_LO_FD} reports examples of Feynman diagrams for each of these contributions.
It is worth pointing out that parton-shower matched predictions contain at the matrix-
element level, among these contributions, only the \textit{Prod.} contribution.
Thus, the first computation of the \textit{Mix} and \textit{Decay} contributions at the matrix-element
level is here reported.
It is interesting to investigate the impact of the latter contributions on the overall
cross section, and, in the following, a comparison of the \textit{Full} result to \textit{Prod.}
will be presented.

At NLO the situation is more involved because of the additional correction terms.
However, the three kind of contributions can still be defined, and the NLO cross section reads
\begin{equation}
\begin{aligned}
& d\sigma^{\text{NLO}}_{\text{Full}} = \bigl( \Gamma_{t,\text{NWA}}^{\text{NLO}}\bigr )^{-2}  \times \\
&
\begin{gathered}
	\Bigl [
	\overbrace{ 
	( d\sigma^{\text{LO}}_{t\bar{t}jj} + d\sigma^{\text{virt}}_{t\bar{t}jj} + d\sigma^{\text{real}}_{t\bar{t}jjj} )
	d\Gamma_{t\bar{t}}^{\text{LO}}}^{\text{Prod.}} + 
	\overbrace{
	d\sigma^{\text{LO}}_{t\bar{t}}
	( d\Gamma_{t\bar{t}jj}^{\text{LO}} + d\Gamma_{t\bar{t}jj}^{\text{virt}} + d\Gamma_{t\bar{t}jjj}^{\text{real}} ) }^{\text{Decay}} + \\
	\left.
	\begin{gathered}
		d\sigma^{\text{LO}}_{t\bar{t}j} d\Gamma_{t\bar{t}j}^{\text{LO}} + d\sigma^{\text{LO}}_{t\bar{t}jj} d\Gamma_{t\bar{t}}^{\text{virt}}
		+ d\sigma^{\text{virt}}_{t\bar{t}} d\Gamma_{t\bar{t}jj}^{\text{LO}} + d\sigma^{\text{virt}}_{t\bar{t}j} d\Gamma_{t\bar{t}j}^{\text{LO}} + \\
		d\sigma^{\text{LO}}_{t\bar{t}j} d\Gamma_{t\bar{t}j}^{\text{virt}} + d\sigma^{\text{real}}_{t\bar{t}jj} d\Gamma_{t\bar{t}j}^{\text{real}}
		+ d\sigma^{\text{real}}_{t\bar{t}j} d\Gamma_{t\bar{t}jj}^{\text{real}}
		\Bigr ]
	\end{gathered} \hspace{5mm}
	\right\} \text{\scriptsize{Mix}}
\end{gathered}
\end{aligned}
\end{equation}
where $\Gamma_{t,\text{NWA}}^{\text{NLO}}$ is the NLO top-quark decay width computed in the NWA limit.
The predictions have been obtained using the setup used by the CMS collaboration in Ref.~\cite{CMS:2022uae}.
However, the cut on the angular separation between the light jet and the $b$
jet $\Delta R(jb) > 0.8$ suppresses the jet activity in the top-quark decays.
Since one of the main goals of this study is to investigate the \textit{Mix} and
\textit{Decay} contributions, predictions have been obtained also for the more
inclusive cut $\Delta R(jb) > 0.4$.
The calculation is performed using a dynamical scale definition and both the
renormalization scale and factorization scale are set to $\mu_R = \mu_F = \mu_0 = H_T/2$,
with
\begin{equation}
H_T = \sum_{i=1}^2 p_T(\ell_i) + \sum_{i=1}^2 p_T(j_i) + \sum_{i=1}^2 p_T(b_i) + p_T^{\text{miss}}.
\end{equation}
The theoretical uncertainty is then estimated performing the $7$-point scale variation.
The NNPDF3.1 PDF set is employed.
Tab.~\ref{tab:ttjj:xs_contr_dr08} reports the predictions for the integrated fiducial
cross section when the cut $\Delta R(jb) > 0.8$ is imposed.
\begin{table}[]
\centering
{\def\arraystretch{1.5}
\begin{tabular}{cccccc}
\hline \hline
 & $\sigma_i^{\text{LO}}$ [fb] &$\sigma_i/\sigma_{\text{Full}}$ & $\sigma_i^{\text{NLO}}$ [fb] & $\sigma_i/\sigma_{\text{Full}}$ & $\mathcal{K}=\sigma^{\text{NLO}}/\sigma^{\text{LO}}$  \\ \hline \hline
\textit{Full} & $868.8(2)^{+60\%}_{-35\%}$ & - & $1225(1)^{+1\%}_{-14\%}$ & - & 1.41 \\
\textit{Prod.} & $843.2(2)^{+60\%}_{-35\%}$ & 0.97 & $1462(1)^{+12\%}_{-19\%}$ & 1.19 & 1.73 \\
\textit{Mix} & $25.465(5)$ & 0.029 & $-236(1)$ & -0.19 & -9.27 \\
\textit{Decay} & $0.2099(1)$ & 0.0002 & $0.1840(8)$ & 0.0002 & 0.88 \\ \hline \hline
\end{tabular}
}
\caption{\textit{Integrated fiducial cross section for the $pp \rightarrow t\bar{t}jj$ process. Results are given for the cut $\Delta R(jb) > 0.8$ with $\mu_0 = H_T/2$ and using NNPDF3.1 PDF set. The various contributions to the Full prediction are also reported. The MC error is reported in parenthesis and, only for Full and Prod., the theoretical uncertainty from scale variation is reported.}}
\label{tab:ttjj:xs_contr_dr08}
\end{table}
These results show that the NLO QCD corrections are of medium size ($41\%$) and substantially
reduce the theoretical uncertainty.
The full prediction is slightly more accurate than the \textit{Prod.}-only prediction:
the inclusion of the \textit{Mix} and \textit{Decay} contributions reduces the theoretical
uncertainty by $5\%$.
At LO the \textit{Prod.} contribution is clearly dominant, with the \textit{Mix} contribution
being only $3\%$ of the total cross section, thus, negligible compared to the theoretical uncertainty.
The \textit{Decay} contribution can be safely disregarded, too, being imperceptible.
On the other hand, at NLO the \textit{Mix} contribution is enhanced in absolute value,
representing $-19\%$ of the total cross section and being now comparable to the theoretical
uncertainty.
Therefore, this contribution cannot be neglected anymore.
The results for $\Delta R(jb) > 0.4$, reported in Tab.~\ref{tab:ttjj:xs_contr_dr04},
\begin{table}[]
\centering
{\def\arraystretch{1.5}
\begin{tabular}{cccccc}
\hline \hline
 & $\sigma_i^{\text{LO}}$ [fb] &$\sigma_i/\sigma_{\text{Full}}$ & $\sigma_i^{\text{NLO}}$ [fb] & $\sigma_i/\sigma_{\text{Full}}$ & $\mathcal{K}=\sigma^{\text{NLO}}/\sigma^{\text{LO}}$  \\ \hline \hline
\textit{Full} & $1074.5(3)^{+60\%}_{-35\%}$ & - & $1460(1)^{+1\%}_{-13\%}$ & - & 1.36 \\
\textit{Prod.} & $983.1(3)^{+60\%}_{-35\%}$ & 0.91 & $1662(1)^{+11\%}_{-18\%}$ & 1.14 & 1.69 \\
\textit{Mix} & $89.42(3)$ & 0.083 & $-205(1)$ & -0.14 & -2.30 \\
\textit{Decay} & $1.909(1)$ & 0.002 & $2.436(6)$ & 0.002 & 1.28 \\ \hline \hline
\end{tabular}
}
\caption{\textit{Integrated fiducial cross section for the $pp \rightarrow t\bar{t}jj$ process. Results are given for the cut $\Delta R(jb) > 0.4$ with $\mu_0 = H_T/2$ and using NNPDF3.1 PDF set. The various contributions to the Full prediction are also reported. The MC error is reported in parenthesis and, only for Full and Prod., the theoretical uncertainty from scale variation is reported.}}
\label{tab:ttjj:xs_contr_dr04}
\end{table}
are very similar, with the \textit{Mix} contribution now larger at LO ($8\%$), but still
negligible compared to the theoretical uncertainty. Once again, at NLO the \textit{Mix}
contribution ($-14\%$) is comparable to the theoretical uncertainty.

\begin{figure}[]
\centering
\begin{subfigure}{0.49\textwidth}
  \includegraphics[width=\linewidth]{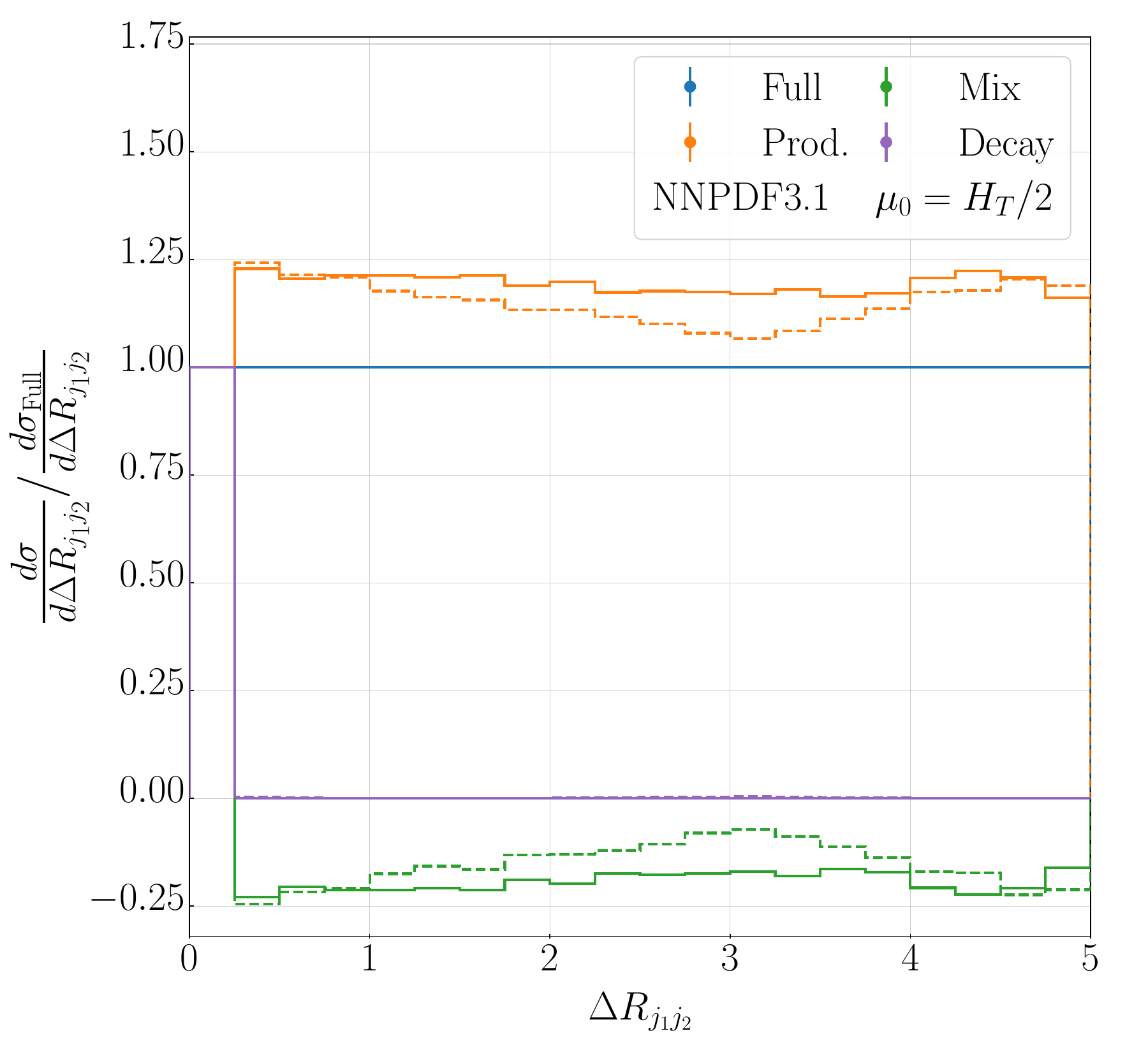}
  \caption*{(a) \textit{Relative size of Prod., Mix and Decay to \textit{Full} for $\Delta R(jb) > 0.8$ (solid line) and $\Delta R(jb) > 0.4$ (dashed line).
         Figure taken from Ref.~\cite{Bevilacqua:2022ozv}.}}
\end{subfigure}
\hfill
\begin{subfigure}{0.49\textwidth}
  \includegraphics[width=\linewidth]{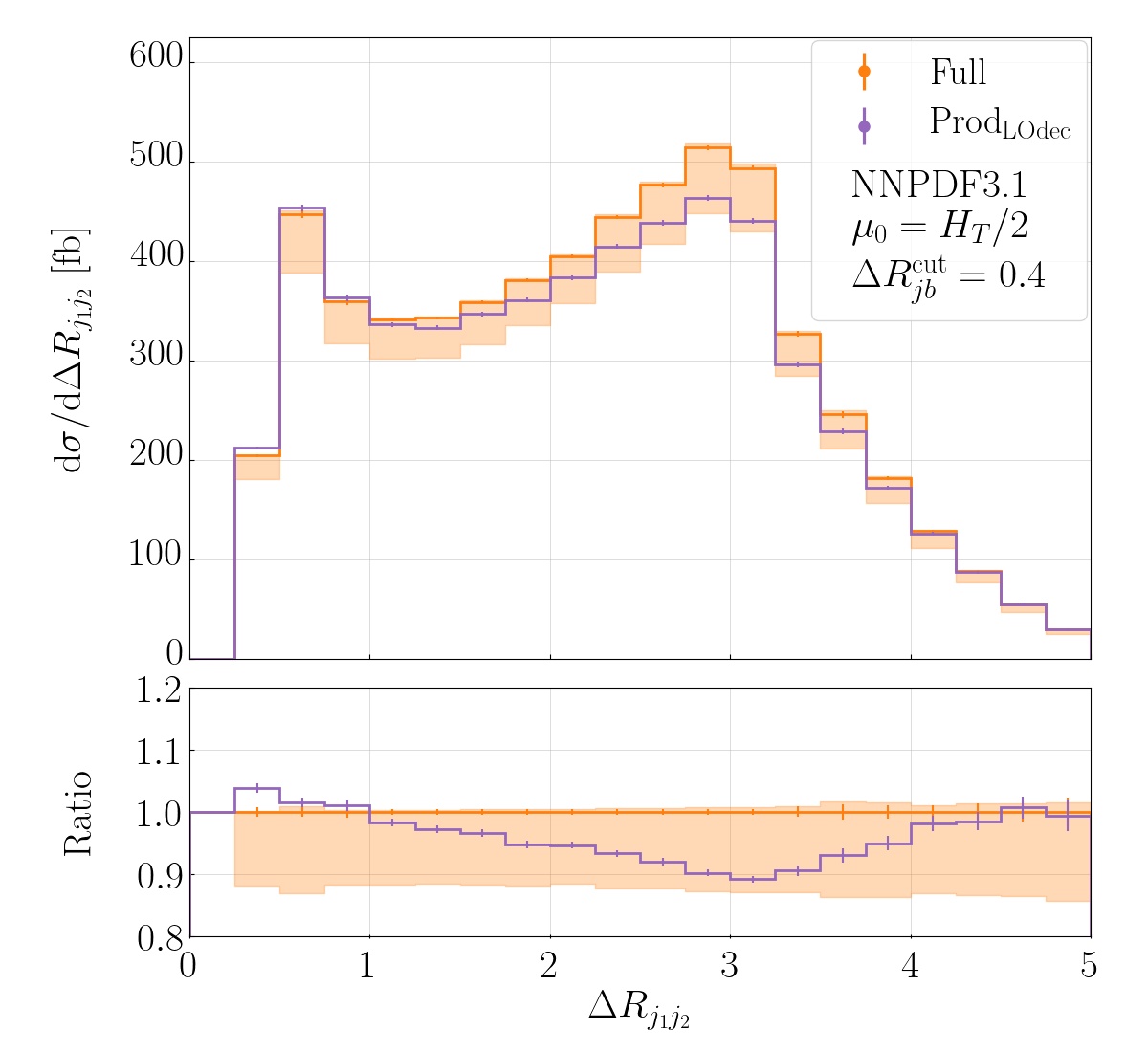}
  \caption*{(b) \textit{Comparison of \textit{Full} and \textit{Prod.} (LOdec) for $\Delta R(jb) > 0.4$. The theoretical uncertainty on \textit{Full} is displayed, too. Figure taken from Ref.~\cite{Lupattelli:2023fgp}.}}
\end{subfigure}
\caption{\textit{Differential cross section distributions as a function of $\Delta R_{j_1 j_2}$ at NLO in QCD for the $pp \rightarrow t\bar{t}jj$ process at the LHC with $\sqrt{s} = 13$ TeV. Results are obtained for $\mu_R = \mu_F = \mu_0 = H_T/2$ and NNPDF3.1 PDF set.}}
\label{fig:ttjj_dr}
\end{figure}

Very interesting is also the behavior of the \textit{Mix} contribution
at the differential level.
Reported in Fig.~\ref{fig:ttjj_dr}a is a plot for the differential fiducial cross section as a function
of the angular separation between the two hardest light jets $\Delta R_{j_1 j_2}$,
normalized to the \textit{Full} prediction.
The solid lines depict predictions where the cut $\Delta R(j b) > 0.8$ is employed,
while the dashed lines illustrate predictions where the more inclusive $\Delta R(j b) > 0.4$
is used.
The \textit{Full} prediction is displayed in blue, the \textit{Prod.} contribution in orange,
the \textit{Mix} contribution in green and the \textit{Decay} contribution in purple.
Once again, we can disregard the \textit{Decay} contribution.
The \textit{Mix} contribution, on the other hand, is quite important and can reach $25\%$.
What is most interesting is the non-trivial kinematic dependence of \textit{Prod.} and
\textit{Mix} when the $\Delta R(j b) > 0.4$ cut is used.
Thus, it is interesting to see how this kinematic dependence affects the absolute
$\Delta R_{j_1 j_2}$ distribution.

Before delving into that, we need to define the $Prod_{LOdec}$ cross section. This is the NLO cross section with LO
top-quark decays and where the additional light jets are emitted only in the production stage of the
top-quark pair.
This represents the actual fixed-order part of parton-shower calculations and is a mere rescaling
of the \textit{Prod.} contribution by a factor
$(\Gamma_{t,\text{NWA}}^{\text{NLO}}/\Gamma_{t,\text{NWA}}^{\text{LO}})^2$, i.e. the LO top-quark
width is used instead of the NLO one.

A comparison of the \textit{Full} prediction against the $Prod_{LOdec}$ is displayed in Fig.~\ref{fig:ttjj_dr}b.
The \textit{Mix} contribution clearly affects the shape of this and several other observables
(see Ref.~\cite{Bevilacqua:2022ozv}).
The differences between \textit{Full} and $Prod_{LOdec}$ vary in the range $[-11\%,+4\%]$.
Thus, the shape distortion is $15\%$.
Moreover, $Prod_{LOdec}$ lies outside the uncertainty band of the \textit{Full} prediction in few bins.
Finally, it has been observed that the inclusion of \textit{Mix} slightly reduces the theoretical
uncertainty: for the \textit{Full} prediction, the uncertainty varies in the range $[12\%,14\%]$,
while for the $Prod_{LOdec}$ in the range $[17\%,19\%]$.

\section{Summary}
In this contribution we reported on the calculation of the
$p p \rightarrow \ell^+ \nu_e \ell^- \bar{\nu}_\mu b \bar{b} j j + X$
process at NLO in QCD where, for the first time, the additional light
jets have been consistently included both in the production and decay
stages of the top-quark pair.
The NLO QCD corrections for this process are of medium size ($41\%$
when $\Delta R(jb)>0.8$ is employed and $36\%$ when $\Delta R(jb)>0.4$
is used).
An important result of this study is that the \textit{Mix} contribution
cannot be neglected at NLO, since its size is comparable to the
theoretical uncertainty and it impacts the shape of the differential
distribution.
Therefore, at fixed order, the inclusion of the \textit{Mix} contribution
is imperative.
A comparison of this predictions with results obtained using a parton
shower would be beneficial to asses how well the latter can approximate
the \textit{Mix} contribution.

\Acknowledgements
The research of M. Lupattelli was supported by the  Deutsche Forschungsgemeinschaft 
(DFG) under grant 400140256 - GRK 2497:
\textit{The physics of the heaviest particles at the Large Hadron Collider}
and by the DFG under grant 396021762 - TRR 257:
\textit{P3H - Particle Physics Phenomenology after the Higgs Discovery}.

\bibliography{ttjj_TOP23}{}
\bibliographystyle{unsrt}
 
\end{document}

%% file: econfmacros.tex
\def\beq{\begin{equation}}
\def\eeq#1{\label{#1}\end{equation}}
\def\eeqn{\end{equation}}

\def\beqa{\begin{eqnarray}}
\def\eeqa#1{\label{#1}\end{eqnarray}}
\def\eeqan{\end{eqnarray}}

\let\bar=\overbar

\def\Dslash{\not{\hbox{\kern-4pt $D$}}}
\def\dslash{\not{\hbox{\kern-2pt $\del$}}}

\def\msb{{\bar{\ssstyle M \kern -1pt S}}}

